\documentclass{ws-procs9x6}
\usepackage{ws-toc,ws-multind}
\usepackage{color}
\makeindex{author}                     

\begin{document}



\cleardoublepage




\wstoc{Generalized Keller-Segel models of chemotaxis. Analogy with nonlinear mean field Fokker-Planck equations}{P.H. Chavanis}

\title{GENERALIZED KELLER-SEGEL MODELS OF CHEMOTAXIS. ANALOGY WITH NONLINEAR MEAN FIELD FOKKER-PLANCK EQUATIONS}

\author{PIERRE-HENRI CHAVANIS$^\ddag$}
\aindx{Chavanis, P.H.}

\address{ Laboratoire de Physique Th\'eorique, Universit\'e Paul
Sabatier, 118 route de Narbonne 31062 Toulouse, France\\
\email{$^\ddag\!$E-mail: chavanis@irsamc.ups-tlse.fr}}

   \begin{abstract}We consider a generalized class of Keller-Segel
   models describing the chemotaxis of biological populations
   (bacteria, amoebae, endothelial cells, social insects,...). We
   show the analogy with nonlinear mean field Fokker-Planck equations
   and generalized thermodynamics.  As an illustration, we introduce a new
   model of chemotaxis incorporating both effects of anomalous
   diffusion and exclusion principle (volume filling). We also discuss
   the analogy between biological populations described by the
   Keller-Segel model and self-gravitating Brownian particles
   described by the Smoluchowski-Poisson system.

\end{abstract}

\bodymatter
  
%

\section{Introduction}
\label{sec_introduction}

The name chemotaxis refers to the motion of organisms induced by
chemical signals \cite{murray}. In some cases, the biological
organisms (bacteria, amoebae, endothelial cells, ants...) secrete a
substance (pheromone, smell, food, ...) that has an attractive effect
on the organisms themselves. Therefore, in addition to their diffusive
motion, they move preferentially along the gradient of concentration
of the chemical they secrete (chemotactic flux). When attraction
prevails over diffusion, the chemotaxis can trigger a
self-accelerating process until a point at which aggregation takes
place. This is the case for the slime mold {\it Dictyostelium
discoideum} and for the bacteria {\it Escherichia coli}. This is
referred to as chemotactic collapse. A model of slime mold aggregation
has been introduced by Patlak \cite{patlak} and Keller \& Segel
\cite{ks} in the form of two coupled differential equations.
The first equation is a drift-diffusion equation describing the
evolution of the concentration of bacteria and the second equation is
a diffusion equation with terms of source and degradation describing
the evolution of the concentration of the chemical. In the simplest
model, the diffusion coefficient $D$ and the mobility $\chi$ of the
bacteria are constant. This forms the standard Keller-Segel
model. However, the original Keller-Segel model allows these
coefficients to depend on the concentration of the bacteria and of the
chemical. If we assume that they only depend on the concentration of
the bacteria, the general Keller-Segel model becomes similar to a
nonlinear mean field Fokker-Planck equation. Nonlinear Fokker-Planck
(NFP) equations have been introduced in a very different context, in
relation with a notion of generalized thermodynamics \cite{frank}. As
far as we know, the connection between the general Keller-Segel model
and nonlinear mean field Fokker-Planck equations has been first
mentioned in Chavanis \cite{gen} and developed in subsequent papers
(see \cite{csbio} and references therein).  This analogy makes
possible to interpret results obtained in chemotaxis in terms of a
generalized thermodynamics. At the same time, chemotaxis becomes an
example of great physical importance for which a notion of (effective)
generalized thermodynamics is justified.

The standard Keller-Segel (KS) model has been extensively studied in the
mathematical literature (see the review by Horstmann
\cite{horstmann}). It was found early that, above a critical mass, the
distribution of bacteria becomes unstable and collapses.  This {\it
chemotactic collapse} leads ultimately to the formation of Dirac peaks
\cite{nanjundiah,childress,jager,nagai,othmer,herrero96,herrero97,herrero98,biler,brenner,dolbeault,biler1,biler2}. Recently, it was shown by Chavanis, Rosier \& Sire \cite{crs} that, when the equation for the evolution of the concentration is approximated by a Poisson equation \cite{jager,herrero97,brenner}, the standard Keller-Segel (KS) model is isomorphic to the Smoluchowski-Poisson (SP) system 
describing self-gravitating Brownian particles. The chemotactic
collapse of biological populations above a critical mass is equivalent
to the {\it gravitational collapse} of self-gravitating Brownian
particles below a critical temperature
\cite{mass}. Assuming that the evolution is spherically symmetric,
Chavanis \& Sire \cite{crs,sc,post,tcoll,sopik,virial1,virial2,mass}
were able to describe all the phases of the collapse (pre-collapse and
post-collapse) {\it analytically} in $d$ dimensions, including the
critical dimension $d=2$.

Recently, some authors have considered generalizations of the standard
Keller-Segel (KS) model. Two main classes of generalized Keller-Segel (GKS)
models of chemotaxis have been introduced:

{\it (i) Models with filling factor:} Hillen \& Painter
\cite{hillen,painter} considered a model with a normal diffusion and a
density-dependent mobility $\chi(\rho)=\chi(1-\rho/\sigma_{0})$
vanishing above a maximum density $\sigma_{0}$. The same model was
introduced independently by Chavanis \cite{gen,degrad} in relation
with an ``exclusion principle'' connected to the Fermi-Dirac entropy
in physical space.  In these models, the density of bacteria remains
always bounded by the maximum density: $\rho({\bf r},t)\le\sigma_{0}$.
This takes into account finite size effects and filling
factors. Indeed, since the cells have a finite size, they cannot be
compressed indefinitely. In this generalized Keller-Segel model,
chemotactic collapse leads ultimately to the formation of a {\it
smooth aggregate} instead of a {\it Dirac peak} in the standard
Keller-Segel model. This regularized model prevents finite time-blow
up and the formation of (unphysical) singularities like infinite
density profiles and Dirac peaks.  Therefore, the Dirac peaks
(singularities) are replaced by smooth density profiles (aggregates).

{\it (ii) Models with anomalous diffusion:} Chavanis \& Sire
\cite{lang} studied a model with a constant mobility and a power law
diffusion coefficient $D(\rho)=D\rho^{\gamma-1}$ (with
$\gamma=1+1/n$). This lead to a process of anomalous diffusion
connected to the Tsallis entropy \cite{tsallis}. For
$0<n<n_{3}=d/(d-2)$, the system reaches a self-confined distribution
similar to a stable polytrope (e.g. a classical white dwarf star) in
astrophysics. For $n>n_{3}$, the system undergoes chemotactic collapse
above a critical mass (the classical chemotactic collapse related to
the standard Keller-Segel model is recovered for $n\rightarrow
+\infty$) \cite{lang}. In the pre-collapse regime, the evolution is
self-similar and leads to a finite time singularity. A Dirac peak is
formed in the post-collapse regime. For $n=n_{3}$, the dynamics is
peculiar and involves a critical mass similar to the Chandrasekhar
limiting mass of relativistic white dwarf stars in astrophysics
\cite{csmass}. The case of negative index $n<0$ is treated in
\cite{logotrope} with particular emphasis on the index $n=-1$ leading
to {\it logotropes}.

In the present paper, we discuss a larger class of generalized
Keller-Segel models and interprete these equations in relation with
nonlinear mean field Fokker-Planck equations and generalized
thermodynamics. For illustration, we present for the first time a
model incorpoating both a filling factor and some effects of anomalous
diffusion.

\section{The generalized Keller-Segel model}
\label{sec_reg}

\subsection{The dynamical equations}
\label{sec_dyn}

The general Keller-Segel model \cite{ks} describing the chemotaxis of bacterial populations consists in two coupled differential equations 
\begin{eqnarray}
\label{dyn1}
\frac{\partial\rho}{\partial t}=\nabla\cdot \left (D_{2}\nabla\rho)-\nabla\cdot (D_{1}\nabla c\right ),
\end{eqnarray}
\begin{eqnarray}
\label{dyn2}
\epsilon\frac{\partial c}{\partial t}=-k(c)c+f(c)\rho+D_{c}\Delta c, 
\end{eqnarray}
that govern the evolution of the density of bacteria $\rho({\bf r},t)$
and the evolution of the secreted chemical $c({\bf r},t)$. The
bacteria diffuse with a diffusion coefficient $D_{2}$ and they also
move in a direction of a positive gradient of the chemical
(chemotactic drift). The coefficient $D_{1}$ is a measure of the
strength of the influence of the chemical gradient on the flow of
bacteria. On the other hand, the chemical is produced by the bacteria
with a rate $f(c)$ and is degraded with a rate $k(c)$. It also
diffuses with a diffusion coefficient $D_{c}$. In the general
Keller-Segel model, $D_{1}=D_{1}(\rho,c)$ and $D_{2}=D_{2}(\rho,c)$
can both depend on the concentration of the bacteria and of the
chemical. This takes into account microscopic constraints, like
close-packing effects, that can hinder the movement of bacteria. If we
assume that the quantities only depend on the concentration of
bacteria and write $D_{2}=Dh(\rho)$, $D_{1}=\chi g(\rho)$,
$k(c)=k^{2}$, $f(c)=\lambda$ and $D_{c}=1$, we obtain
\begin{eqnarray}
\label{dyn3}
\frac{\partial\rho}{\partial t}=\nabla\cdot \left ( D h(\rho)\nabla\rho-
\chi g(\rho) \nabla c\right ),
\end{eqnarray}
\begin{eqnarray}
\label{dyn4}
\epsilon\frac{\partial c}{\partial t}=\Delta c-k^{2}c+\lambda\rho. 
\end{eqnarray}
For $\epsilon=0$, Eq. (\ref{dyn4}) becomes the screened Poisson equation
\begin{eqnarray}
\label{dyn5}
\Delta c-k^{2}c=-\lambda\rho. 
\end{eqnarray}
Therefore, we can identify $k^{-1}$ as the screening length. If we
assume furthermore that $k=0$, we get the Poisson equation
\begin{eqnarray}
\label{dyn6}
\Delta c=-\lambda\rho. 
\end{eqnarray}

The generalized Keller-Segel (GKS) model (\ref{dyn3}) can be viewed as
a nonlinear mean-field Fokker-Planck (NFP) equation
\cite{gen}. Written in the form $\partial_t\rho=\nabla\cdot (\nabla
(D(\rho)\rho)-\chi(\rho)\rho\nabla c)$, it is associated with a
stochastic Ito-Langevin equation
\begin{eqnarray}
\label{dyn7}
\frac{d{\bf r}}{dt}=\chi(\rho)\nabla c+\sqrt{2D(\rho)}{\bf R}(t),
\end{eqnarray}
with
\begin{eqnarray}
\label{dyn8}
\chi(\rho)=\frac{\chi g(\rho)}{\rho}, \quad D(\rho)=\frac{D}{\rho}\int_{0}^{\rho}h(x)dx,
\end{eqnarray} 
where ${\bf R}(t)$ is a white noise satisfying $\langle {\bf
R}(t)\rangle={\bf 0}$ and $\langle R_{i}(t)R_{j}(t')\rangle
=\delta_{ij}\delta(t-t')$ where $i=1,...,d$ label the coordinates of
space. The standard Keller-Segel model is obtained when the mobility
$\chi$ and the diffusion coefficient $D$ are constant. This
corresponds to $h(\rho)=1$ and $g(\rho)=\rho$. In that case, the
stochastic process (\ref{dyn7}) and the Fokker-Planck equation
(\ref{dyn3}) are similar to the ordinary Langevin and Smoluchowski
equations describing the diffusion of a system of particles in a
potential $\Phi({\bf r},t)=-c({\bf r},t)$ that they produce themselves
through Eq. (\ref{dyn4}). For example, when Eq. (\ref{dyn4}) is
approximated by Eq. (\ref{dyn6}), the system becomes isomorphic to the
Smoluchowski-Poisson system describing self-gravitating Brownian
particles \cite{crs,mass}. The steady state of the standard
Keller-Segel equation is $\rho\sim e^{\frac{\chi}{D}c}$. This is
similar to the Boltzmann distribution $\rho\sim e^{-\Phi/T}$ of
statistical equilibrium provided that we introduce an effective
temperature $T$ through the Einstein relation $T=D/\chi$. In the
present study, we shall consider more general situations and allow the
mobility $\chi(\rho)$ and the diffusion coefficient $D(\rho)$ to
depend on the local concentration of particles $\rho({\bf r},t)$. This
is an heuristic approach to take into account microscopic constraints
that affect the dynamics of the particles at small scales and lead to
non-Boltzmannian distributions at equilibrium. Indeed, it is not
surprising that the mobility or the diffusive properties of a particle
depend on its environement. For example, in a dense medium its motion
can be hampered by the presence of the other particles so that its
mobility is reduced.

\subsection{Generalized free energy and H-theorem}
\label{sec_h}

We define the energy by 
\begin{eqnarray}
\label{h3}
E=\frac{1}{2\lambda}\int \left\lbrack (\nabla c)^{2}+k^{2} c^{2}\right \rbrack
\, d{\bf r}-\int \rho c \, d{\bf r}.
\end{eqnarray} 
For $\epsilon=0$, this expression  reduces to 
\begin{eqnarray}
\label{h2}
E=-\frac{1}{2}\int \rho c \, d{\bf r}.
\end{eqnarray} 
On the other hand, we define the temperature by
\begin{eqnarray}
\label{h4}
T=\frac{D}{\chi}.
\end{eqnarray} 
Therefore, the Einstein relation is preserved in the generalized
thermodynamical framework. We also set $\beta=1/T$. We introduce the
generalized entropic functional
\begin{eqnarray}
\label{h5}
S=-\int C(\rho)\, d{\bf r},
\end{eqnarray}
where $C(\rho)$ is a convex function ($C''\ge 0$) defined by
\begin{eqnarray}
\label{h6}
C''(\rho)=\frac{h(\rho)}{g(\rho)}.
\end{eqnarray}
This defines the entropy up to a term of the form $AM+B$ where
$M=\int\rho d{\bf r}$ is the mass (which is a conserved quantity). We
can adapt the values of the constants $A$ and $B$ in order to obtain
convenient expressions of the entropy. Finally, we introduce the
generalized free energy
\begin{eqnarray}
\label{h7}
F=E-TS.
\end{eqnarray}
The definition of the free energy (Legendre transform) is preserved in
the generalized thermodynamical framework. The free energy is the correct 
thermodynamical potential since the system is {\it dissipative}. Thus, it must
be treated within the canonical ensemble \cite{paper1,gen}.

A straightforward calculation shows that
\begin{eqnarray}
\label{h9}
\dot F=-\frac{1}{\lambda\epsilon}\int (-\Delta c+k^{2}c-\lambda\rho)^{2} d{\bf r}
-\int \frac{1}{\chi g(\rho)}(Dh(\rho)\nabla\rho-\chi g(\rho)\nabla c)^{2}d{\bf r}.\nonumber\\
\end{eqnarray}
For $\epsilon=0$, this equation reduces to
\begin{eqnarray}
\label{h8}
\dot F=-\int \frac{1}{\chi g(\rho)}(Dh(\rho)\nabla\rho-\chi g(\rho)\nabla c)^{2}d{\bf r}.
\end{eqnarray}
Therefore, $\dot F\le 0$ (in all the paper, we assume that
$\epsilon,\lambda,\chi,D,g,h$ are positive quantities). This forms an
$H$ theorem in the canonical ensemble \cite{paper1,gen} for the
nonlinear mean field Fokker-Planck equation (\ref{dyn3}). This also
shows that the free energy $F[\rho,c]$ is the Lyapunov functional of
the generalized Keller-Segel model (\ref{dyn3})-(\ref{dyn4}). It is
sometimes useful to introduce the Massieu function
\begin{eqnarray}
\label{h10}
J=S-\beta E,
\end{eqnarray}
which is related to the free energy by $J=-\beta F$. Clearly, we have
$\dot J\ge 0$. We can now consider particular cases: if $D=0$ (leading
to $T=0$), we get $F=E$ so that $\dot E\le 0$. If $\chi=0$ (leading to
$\beta=0$), we have $J=S$ so that $\dot S\ge 0$.

\subsection{Stationary solution}
\label{sec_sta}

The steady state of Eq. (\ref{dyn3}) satisfies $\dot F=0$. According to Eq. (\ref{h9}), this leads to
\begin{eqnarray}
\label{sta1}
\Delta c-k^{2}c=-\lambda\rho, \qquad Dh(\rho)\nabla\rho-\chi g(\rho)\nabla c={\bf 0}.
\end{eqnarray}
Using Eqs. (\ref{h4}) and (\ref{h6}), the second equation can be rewritten
\begin{eqnarray}
\label{sta3}
C''(\rho)\nabla\rho-\beta\nabla c={\bf 0},
\end{eqnarray}
which can be integrated into
\begin{eqnarray}
\label{sta4}
C'(\rho)=\beta c-\alpha,
\end{eqnarray}
where $\alpha$ is a constant of integration.  Since $C$ is convex, this
equation can be reversed to give
\begin{eqnarray}
\label{sta5}
\rho({\bf r})=F(-\beta c({\bf r})+\alpha),
\end{eqnarray}
where $F(x)=(C')^{-1}(-x)$ is a monotonically decreasing
function. Thus, in the steady state, the density is a monotonically
increasing function $\rho=\rho(c)$ of the concentration.  We have the identity
\begin{eqnarray}
\label{sta6}
\rho'(c)=\frac{\beta}{C''(\rho)}.
\end{eqnarray}
Substituting Eq. (\ref{sta5}) in Eq. (\ref{dyn5}), valid for a
stationary state, we obtain a mean-field equation of the form
\begin{eqnarray}
\label{sta7}
-\Delta c+k^{2}c=\lambda F(-\beta c+\alpha).
\end{eqnarray}
The constant of integration $\alpha$ is determined by the total mass
$M$ (which is a conserved quantity). Finally, we note that the
generalized entropy (\ref{h5}) is related to the distribution
(\ref{sta5}) by:
\begin{eqnarray}
\label{sta8}
C(\rho)=-\int^{\rho}F^{-1}(x)dx.
\end{eqnarray}
Equation (\ref{sta5}) determines the distribution $\rho({\bf r})$ from
the entropy $S$ and Eq. (\ref{sta8}) determines the entropy from the
density.

\subsection{Minimum of free energy}
\label{sec_min}

The critical points of free energy at fixed mass are determined by the variational problem
\begin{eqnarray}
\label{min1}
\delta F+T\alpha\delta M=0,
\end{eqnarray}
where $\alpha$ is a Lagrange multiplier. We can easily establish that
\begin{eqnarray}
\label{min2}
\delta E=-\frac{1}{\lambda}\int (\Delta c-k^{2}c+\lambda\rho)\delta c \, d{\bf r}-\int c\delta\rho \, d{\bf r},
\end{eqnarray}
\begin{eqnarray}
\label{min3}
\delta S=-\int C'(\rho)\delta\rho  \, d{\bf r}.
\end{eqnarray}
The variational problem (\ref{min1}) then leads to 
\begin{eqnarray}
\label{min4}
\Delta c-k^{2}c=-\lambda\rho, \qquad C'(\rho)=\beta c-\alpha.
\end{eqnarray}
Comparing with Eq. (\ref{sta4}), we find that a stationary
solution of Eq. (\ref{dyn3}) is a critical point of $F$ at fixed
mass. On the other hand, we have established that
\begin{eqnarray}
\label{min5}
\dot F\le 0, \qquad \dot F=0 \Leftrightarrow \partial_{t}\rho=0.
\end{eqnarray}
According to Lyapunov's direct method \cite{frank}, this implies that
$\rho({\bf r})$ is linearly dynamically stable with respect to the NFP
equation (\ref{dyn3})-(\ref{dyn4}) iff it is a (local) minimum of $F$
at fixed mass. Maxima or saddle points of $F$ are dynamically
unstable. In conclusion, a steady solution of the GKS model/NFP
equation (\ref{dyn3})-(\ref{dyn4}) is linearly dynamically stable iff
it satisfies (at least locally) the minimization problem:
\begin{eqnarray}
\label{min6}
\min_{\rho,c} \quad \lbrace F[\rho,c]\quad |\quad M[\rho]=M\rbrace.
\end{eqnarray}
 In this sense, dynamical and generalized
thermodynamical stability in the canonical ensemble coincide.
 Furthermore, if $F$ is bounded from below
\footnote{We note that for the standard Keller-Segel model, or for the Smoluchowski-Poisson system, the free 
energy is not bounded from below. In that case, the system can either
relax towards a local minimum of $F$ at fixed mass (when it exists) or
collapse to a Dirac peak \cite{post}, leading to a divergence of the
free energy $F(t)\rightarrow -\infty$. The selection depends on a
complicated basin of attraction. The same situation (basin of
attraction) happens when there exists several minima of free energy at
fixed mass.  }, we can conclude from Lyapunov's theory that the system
will converge to a stable steady state of the GKS model for
$t\rightarrow +\infty$.

Finally, we note that the GKS model can be written 
\begin{eqnarray}
\label{min7}
\frac{\partial\rho}{\partial t}=\nabla\cdot \left \lbrack \chi g(\rho) \nabla\frac{\delta F}{\delta\rho}\right \rbrack,
\end{eqnarray}
where $\delta/\delta\rho$ is the functional derivative. This shows
that the diffusion current ${\bf J}=-\chi g(\rho) \nabla({\delta
F}/{\delta\rho})$ is proportional to the gradient of a quantity
${\delta F}/{\delta\rho}$ that is uniform at equilibrium ($({\delta
F}/{\delta\rho})_{eq}=-T\alpha$ according to Eq. (\ref{min1})). This
corresponds to the linear thermodynamics of Onsager. The same result
can also be obtained from a generalized Maximum Free Energy
Dissipation (MFED) principle which is the variational formulation of
Onsager's linear thermodynamics \cite{gen}.

\subsection{Particular cases}
\label{sec_pc}

If we take $h(\rho)=1$ and  $g(\rho)=1/C''(\rho)$, the NFP equation (\ref{dyn3}) becomes
\begin{eqnarray}
\label{pc1}
\frac{\partial\rho}{\partial t}=\nabla\cdot \left ( D \nabla\rho-
\frac{\chi}{C''(\rho)}  \nabla c\right ).
\end{eqnarray}
In that case, we have a constant diffusion $D(\rho)=D$ and a density
dependent mobility $\chi(\rho)=\chi/(\rho C''(\rho))$. If we take
$g(\rho)=\rho$ and $h(\rho)=\rho C''(\rho)$, the NFP equation
(\ref{dyn3}) becomes
\begin{eqnarray}
\label{pc2}
\frac{\partial\rho}{\partial t}=\nabla\cdot \left ( D\rho C''(\rho) \nabla\rho-{\chi}\rho  \nabla c\right ).
\end{eqnarray}
In that case, we have a constant mobility $\chi(\rho)=\chi$ and a
density dependent diffusion $D(\rho)=D\rho \lbrack
C(\rho)/\rho\rbrack'$. Note that the condition $D(\rho)\ge 0$ requires
that $\lbrack C(\rho)/\rho\rbrack'\ge 0$. This gives a constraint on
the possible forms of $C(\rho)$. 

Finally, if we multiply the diffusion term and the drift term in the
NFP equation (\ref{dyn3}) by the {\it same} positive function
$\lambda({\bf r},t)$ (which can be for example a function of
$\rho({\bf r},t)$), we obtain a NFP equation having the same
free energy (i.e. satisfying an $H$-theorem $\dot F\le 0$) and the same equilibrium states as the original one. Therefore,
for a given entropy $C(\rho)$, we can form an {\it infinite class} of
NFP equations possessing the same general properties \cite{gen}.

\subsection{Generalized Smoluchowski equation}
\label{sec_gs}

The NFP equation (\ref{pc2}) can be written in the form of a generalized
Smoluchowski (GS) equation
\begin{eqnarray}
\label{gs1}
\frac{\partial\rho}{\partial t}=\nabla\cdot \left \lbrack  \chi (\nabla p-
\rho  \nabla c )\right \rbrack,
\end{eqnarray}
with a barotropic equation of state $p(\rho)$ given by
\begin{eqnarray}
\label{gs2}
p'(\rho)=T\rho C''(\rho).
\end{eqnarray}
Since $C$ is convex, we have $p'(\rho)\ge 0$. 
Integrating Eq. (\ref{gs2}) twice, we get
\begin{eqnarray}
\label{gs3}
TC(\rho)=\rho \int^{\rho}\frac{p(\rho')}{\rho^{'2}}d\rho'.
\end{eqnarray}
Therefore, the free energy (\ref{h7}) can be rewritten 
\begin{eqnarray}
\label{gs4}
F=\frac{1}{2\lambda}\int \left\lbrack (\nabla c)^{2}+k^{2} c^{2}\right \rbrack
\, d{\bf r}-\int \rho c \, d{\bf r}+\int  \rho \int^{\rho}\frac{p(\rho')}{\rho^{'2}}d\rho'd{\bf r}.
\end{eqnarray}
With these notations, the $H$-theorem becomes 
\begin{eqnarray}
\label{gs5}
\dot F=-\frac{1}{\lambda\epsilon}\int (\Delta c-k^{2}c+\lambda\rho)^{2} d{\bf r}
-\int \frac{1}{\chi \rho}(\nabla p-\rho\nabla c)^{2}d{\bf r}\le 0.
\end{eqnarray}
The stationary solutions of the GS equation (\ref{gs1}) satisfy the relation
\begin{eqnarray}
\label{gs6}
\nabla p-\rho\nabla c={\bf 0},
\end{eqnarray}
which is similar to a condition of hydrostatic equilibrium. Since $p=p(\rho)$, this relation can be integrated to give $\rho=\rho(c)$ through
\begin{eqnarray}
\label{gs7}
\int^{\rho}\frac{p'(\rho')}{\rho'}d\rho'=c.
\end{eqnarray}
This is equivalent to
\begin{eqnarray}
\label{gs8}
\frac{p'(\rho)}{\rho}=\frac{1}{\rho'(c)}.
\end{eqnarray}
This relation can also be obtained from Eqs. (\ref{gs2}) and
(\ref{sta6}). Therefore, we recover the fact that, in the steady
state, $\rho=\rho(c)$ is a monotonically increasing function of
$c$. We also note the identity
\begin{eqnarray}
\label{gs9}
p(\rho)=\frac{1}{\chi}D(\rho)\rho=T\rho^{2} \left \lbrack \frac{C(\rho)}{\rho} \right\rbrack'=T\lbrack C'(\rho)\rho-C(\rho)\rbrack. 
\end{eqnarray}
Finally, we note that the relation (\ref{gs7}) can also be obtained by
extremizing the free energy (\ref{gs4}) at fixed mass writing $\delta
F-\alpha\delta M=0$. More precisely, we have the important result:
{\it a steady solution of the generalized Smoluchowski equation
(\ref{gs1})-(\ref{dyn4}) is linearly dynamically stable iff it is a
(local) minimum of the free energy $F[\rho,c]$ at fixed mass
$M[\rho]=M$.} This corresponds to the minimization problem
(\ref{min6}).

The generalized Smoluchowski equation (\ref{gs1}) can also be
obtained formally from the damped Euler equations \cite{gen}:
\begin{eqnarray}
\label{damped1} {\partial\rho\over\partial t}+\nabla\cdot (\rho{\bf u})=0,
\end{eqnarray}
\begin{eqnarray}
\label{damped2}\frac{\partial {\bf u}}{\partial t}+({\bf u}\cdot \nabla){\bf u}=-\frac{1}{\rho}\nabla p+\nabla c-\xi {\bf u}.
\end{eqnarray}
For $\xi=0$, we recover the usual barotropic Euler equations of
hydrodynamics. Alternatively, if we consider the strong friction limit
$\xi\rightarrow +\infty$, we can formally neglect the inertial term in
Eq. (\ref{damped2}) and we get $\xi {\bf u}=-\frac{1}{\rho}\nabla
p+\nabla c+O(\xi^{-1})$. Substituting this relation in the continuity
equation (\ref{damped1}), we obtain the generalized Smoluchowski
equation (\ref{gs1}) with $\chi=1/\xi$. These hydrodynamic equations
(hyperbolic model) have been proposed in the context of chemotaxis to
describe the organization of endothelial cells
\cite{gamba,filbet,csbio,jeans}. This inertial model takes into account the
fact that the cells do not respond immediately to the chemotactic
drift but that they have the tendency to continue in a given direction
on their own. Therefore, the inertial term models cells directional
persistence while the general density dependent pressure term $-\nabla
p(\rho)$ takes into account anomalous diffusion or the fact that the
cells do not interpenetrate. Finally, the friction force $-\xi{\bf u}$
measures the response of the system to the chemotactic ``force''
$\nabla c$. Indeed, after a relaxation time of the order $\xi^{-1}$
their velocity will be aligned with the chemotactic gradient.  For
$\xi=0$, Eqs. (\ref{damped1})-(\ref{damped2}) lead to the formation of
filaments that are interpreted as the beginning of a vasculature
\cite{gamba,filbet,csbio,jeans}.  
These filaments, or networks patterns, are not obtained in the
Keller-Segel model (parabolic model), corresponding to $\xi\rightarrow
+\infty$, which leads to point-wise blow up or round aggregates
\cite{horstmann,sc}. Note finally, that the GS equation (\ref{gs1}) can be derived rigorously from kinetic models in a strong friction limit
$\xi\rightarrow +\infty$, using a Chapman-Enskog expansion \cite{cll}
or a method of moments \cite{csbio}.

\subsection{Kinetic derivation of the generalized Keller-Segel model}
\label{sec_kin}

As discussed previously, the generalized Keller-Segel model
(\ref{dyn3}) can be viewed as a nonlinear Fokker-Planck equation where
the diffusion coefficient and the mobility explicitly depend on the
local concentration of particles. Such generalized Fokker-Planck
equations can be derived from a kinetic theory, starting from the
master equation, and assuming that the probabilities of transition
explicitly depend on the occupation number (concentration) of the
initial and arrival states. Below, we briefly summarize and adapt to
the present situation the approach developed by Kaniadakis \cite{k1}
in a more general context.

We introduce a stochastic dynamics by defining the probability of
transition of a particle from position ${\bf r}$ to position ${\bf
r}'$. Following Kaniadakis \cite{k1}, we assume the following
form
\begin{eqnarray}
\label{kin1}
\pi({\bf r}\rightarrow {\bf r}')=w({\bf r},{\bf r}-{\bf r}')a\lbrack\rho({\bf r},t)\rbrack b\lbrack\rho({\bf r}',t)\rbrack.
\end{eqnarray}
Usual stochastic processes correspond to $a(\rho)=\rho$ and
$b(\rho)=1$: the probability of transition is proportional to the
density of the initial state and independent on the density of the
final state.  They lead to the ordinary Fokker-Planck equation
(\ref{sm1}) as will be shown below. Here, we assume a more general
dependence on the occupancy in the initial and arrival states. This
can account for microscopic constraints like close-packing effects
that can inhibitate the transition. Quite generally, the evolution of
the density satisfies the master equation
\begin{eqnarray}
\label{kin2}
\frac{\partial\rho}{\partial t}=\int \left\lbrack \pi({\bf r}'\rightarrow {\bf r})-\pi({\bf r}\rightarrow {\bf r}')\right\rbrack d{\bf r}'.
\end{eqnarray}
Assuming that the evolution is sufficiently slow, and local, such that the dynamics only permits values of ${\bf r}'$ close to ${\bf r}$, one can develop the term in brackets in Eq. (\ref{kin2}) in powers of  ${\bf r}-{\bf r}'$. Proceeding along the lines of \cite{k1}, we obtain a Fokker-Planck-like equation 
\begin{equation}
\label{kin3}
\frac{\partial\rho}{\partial t}=\frac{\partial}{\partial x_{i}}\left\lbrack\left (\zeta_{i}+\frac{\partial\zeta_{ij}}{\partial x_{j}}\right )\gamma(\rho)+\gamma(\rho)\frac{\partial\ln \kappa(\rho)}{\partial\rho}\zeta_{ij}\frac{\partial\rho}{\partial x_{j}}\right\rbrack,
\end{equation}
with
\begin{equation}
\label{kin4}
\gamma(\rho)=a(\rho)b(\rho),\qquad \kappa(\rho)=\frac{a(\rho)}{b(\rho)},
\end{equation}
and
\begin{equation}
\label{kin5}
\zeta_{i}({\bf r})=-\int y_{i}w({\bf r},{\bf y})d{\bf y},
\end{equation}
\begin{equation}
\label{kin6}
\zeta_{ij}({\bf r})=\frac{1}{2}\int y_{i}y_{j}w({\bf r},{\bf y})d{\bf y}.
\end{equation}
The moments $\zeta_{i}$ and $\zeta_{ij}$ are fixed by the Langevin equation 
\begin{eqnarray}
\label{kin7} \frac{d{\bf
r}}{dt}=\chi\nabla c+\sqrt{2D}{\bf R}(t).
\end{eqnarray}
Assuming isotropy $\zeta_{i}=J_{i}$, $\zeta_{ij}=D\delta_{ij}$,
the kinetic equation (\ref{kin3}) becomes
\begin{equation}
\label{kin8}
\frac{\partial\rho}{\partial t}=\nabla\cdot \left\lbrack ({\bf J}+\nabla D)\gamma(\rho)-\gamma(\rho)\frac{\partial\ln \kappa(\rho)}{\partial\rho}D\nabla c\right\rbrack.
\end{equation}
Now, according to the Langevin equation (\ref{kin7}), $D$ is independent on ${\bf r}$ and ${\bf J}=-\chi\nabla c$. Thus, we get
\begin{equation}
\label{kin9}
\frac{\partial\rho}{\partial t}=\nabla\cdot \left\lbrack D\gamma(\rho)\frac{\partial\ln \kappa(\rho)}{\partial\rho}\nabla\rho-\chi\gamma(\rho)\nabla c \right\rbrack.
\end{equation}
If we define
\begin{equation}
\label{kin10}
h(\rho)=\gamma(\rho)\frac{\partial\ln \kappa(\rho)}{\partial\rho}, \qquad g(\rho)=\gamma(\rho),
\end{equation}
the foregoing equation can be rewritten
\begin{eqnarray}
\label{kin11a}
\frac{\partial\rho}{\partial t}=\nabla\cdot \left ( Dh(\rho)\nabla\rho-\chi g(\rho)\nabla c\right ),
\end{eqnarray}
and it coincides with the GKS model (\ref{dyn3}). We note that
\begin{eqnarray}
\label{kin11b}
\ln\kappa(\rho)=C'(\rho).
\end{eqnarray}
We also have the relations
\begin{eqnarray}
\label{kin12}
a(\rho)=\sqrt{\gamma(\rho)\kappa(\rho)}=\sqrt{g(\rho)}e^{C'(\rho)/2},
\end{eqnarray}
\begin{eqnarray}
\label{kin13}
b(\rho)=\sqrt{\frac{\gamma(\rho)}{\kappa(\rho)}}=\sqrt{g(\rho)}e^{-C'(\rho)/2}.
\end{eqnarray}
Inversely,
\begin{eqnarray}
\label{kin14}
g(\rho)=a(\rho)b(\rho), \qquad C'(\rho)=\ln\left\lbrack \frac{a(\rho)}{b(\rho)}\right\rbrack,
\end{eqnarray}
\begin{eqnarray}
\label{kin15}
h(\rho)=b(\rho)a'(\rho)-a(\rho)b'(\rho).
\end{eqnarray}

It seems natural to assume that the transition probability is
proportional to the density of the initial state so that
$a(\rho)=\rho$. In that case, we obtain an equation of the form
\begin{equation}
\label{kin16}
\frac{\partial\rho}{\partial t}=\nabla\cdot \left ( D\left\lbrack b(\rho)-\rho b'(\rho)\right \rbrack \nabla\rho-\chi\rho b(\rho)\nabla c\right ). 
\end{equation}
Note that the coefficients of diffusion and mobility are not
independent since they are both expressed in terms of
$b(\rho)$. Choosing $b(\rho)=1$, i.e. a probability of transition
which does not depend on the population of the arrival state, leads to
the standard Fokker-Planck equation, or standard Keller-Segel model
(\ref{sm1}).  If, now, we assume that the transition probability is
blocked (inhibited) if the concentration of the arrival state is equal
to an upper bound $\sigma_0$, then it seems natural to take
$b(\rho)=1-\rho/\sigma_{0}$. In that case, we obtain
\begin{equation}
\label{kin17}
\frac{\partial\rho}{\partial t}=\nabla\cdot \left ( D\nabla\rho-\chi\rho (1-\rho/\sigma_{0})\nabla c\right ),
\end{equation}
which will be considered in Sec. \ref{sec_ff}. Inversely, we can
wonder what the general form of the mobility will be if we assume a
normal diffusion $h(\rho)=1$. This leads to $b(\rho)-\rho b'(\rho)=1$
which is integrated in $b(\rho)=1+K\rho$ where $K$ is a
constant. Interestingly, we find that this condition selects the class
of fermions ($K=-1$) and bosons ($K=+1$) and intermediate statistics
(arbitrary $K$). The corresponding NFP equation is
\begin{equation}
\label{kin18}
\frac{\partial\rho}{\partial t}=\nabla\cdot \left ( D\nabla\rho-\chi\rho (1+K\rho)\nabla c\right ).
\end{equation}

\section{Examples of generalized Keller-Segel models}
\label{sec_ex}

In this section, we consider generalized Keller-Segel models of
chemotaxis  and show their relation with a
formalism of generalized thermodynamics.

\subsection{The standard Keller-Segel model: Boltzmann entropy}
\label{sec_sm}

If we take $h(\rho)=1$ and $g(\rho)=\rho$, we get the standard Keller-Segel model
\begin{eqnarray}
\label{sm1}
\frac{\partial\rho}{\partial t}=\nabla\cdot \left ( D \nabla\rho-
\chi \rho \nabla c\right ).
\end{eqnarray}
It corresponds to an ordinary diffusion $D(\rho)=D$ and a constant
mobility $\chi(\rho)=\chi$.  The associated stochastic process is the ordinary Langevin equation
\begin{eqnarray}
\label{sm2}
\frac{d{\bf r}}{dt}=\chi\nabla c+\sqrt{2D}{\bf R}(t).
\end{eqnarray}
The entropy is the Boltzmann entropy 
\begin{eqnarray}
\label{sm3}
S=-\int \rho\ln\rho d{\bf r},
\end{eqnarray}
and the stationary solution of Eq. (\ref{sm1}) is the Boltzmann
distribution
\begin{eqnarray}
\label{sm4}
\rho=e^{\beta c-\alpha-1}.
\end{eqnarray}
The standard Keller-Segel model is isomorphic to the Smoluchowski equation 
with an isothermal equation of state  
\begin{eqnarray}
\label{sm5}
p(\rho)=\rho T.
\end{eqnarray}

\subsection{Generalized Keller-Segel model with power law diffusion: Tsallis entropy}
\label{sec_pld}

If we take $h(\rho)=q\rho^{q-1}$ and $g(\rho)=\rho$, we obtain the GKS model
\begin{eqnarray}
\label{pld1}
\frac{\partial\rho}{\partial t}=\nabla\cdot \left ( D \nabla\rho^{q}-
\chi \rho \nabla c\right ).
\end{eqnarray}
It corresponds to a power law diffusion $D(\rho)=D\rho^{q-1}$ and a
constant mobility $\chi(\rho)=\chi$.  The associated stochastic
process is
\begin{eqnarray}
\label{pld2}
\frac{d{\bf r}}{dt}=\chi\nabla c+\sqrt{2D}\rho^{\frac{q-1}{2}}{\bf R}(t).
\end{eqnarray}
This model can take into account effects of non-ergodicity and
nonextensivity.  It leads to a situation of anomalous diffusion
related to the Tsallis statistics \cite{tsallis}.  For $q=1$, we
recover the standard Keller-Segel model with a constant diffusion
coefficient, corresponding to a pure random walk (Brownian model). In
that case, the sizes of the random kicks are uniform and do not depend
on where the particle happens to be. For $q\neq 1$, the size of the
random kicks changes, depending on the distribution of the particles
around the ``test'' particle. A particle which is in a region that is
highly populated [large $\rho({\bf r},t)$] will tend to have larger
kicks if $q>1$ and smaller kicks if $q<1$. Since the microscopics
depends on the actual density in phase space, this creates a bias in
the ergodic behavior of the system.  Then, the dynamics has a fractal
or multi-fractal phase space structure. The generalized entropy
associated to Eq. (\ref{pld1}) is the Tsallis entropy
\begin{eqnarray}
\label{pld3}
S=-\frac{1}{q-1}\int (\rho^{q}-\rho) d{\bf r},
\end{eqnarray}
and the stationary solution is the Tsallis distribution
\begin{eqnarray}
\label{pld4}
\rho=\left (\frac{1}{q}\right )^{\frac{1}{q-1}}\left\lbrack 1-(q-1)(-\beta c+\alpha)\right\rbrack_{+}^{1/(q-1)}.
\end{eqnarray}
The generalized Keller-Segel model (\ref{pld1}) is isomorphic to the generalized Smoluchowski equation (\ref{gs1}) with an equation of state
\begin{eqnarray}
\label{pld5}
p(\rho)=T\rho^{q}.
\end{eqnarray}
This is similar to a polytropic gas with an equation of state
$p=K\rho^{\gamma}$ (with $\gamma=1+1/n$) where $K=T$ plays the role of
a polytropic temperature and $q=\gamma$ is the polytropic index. For
$q=1$, we recover the standard Keller-Segel model (\ref{sm1}). For
$q=2$, we have some simplifications. In that case, the GKS model
(\ref{pld1}) becomes
\begin{eqnarray}
\label{pld10}
\frac{\partial\rho}{\partial t}=\nabla\cdot \left ( D \nabla\rho^{2}-\chi \rho \nabla c\right ).
\end{eqnarray}
The entropy is the quadratic functional
\begin{eqnarray}
\label{pld11}
S=-\int \rho^{2} d{\bf r},
\end{eqnarray}
and the stationary solution is 
\begin{eqnarray}
\label{pld12}
\rho=-\frac{1}{2}(-\beta c+\alpha),
\end{eqnarray}
corresponding to a linear relation between the density and the
concentration. In that case, the differential equation (\ref{sta7})
determining the steady state reduces to the Helmholtz
equation. Finally, the pressure is
\begin{eqnarray}
\label{pld13}
p(\rho)=T\rho^{2}.
\end{eqnarray}
In the context of generalized thermodynamics, the NFP equation
(\ref{pld1}) was introduced by Plastino \& Plastino \cite{pp} and the
generalized stochastic process (\ref{pld2}) was introduced by Borland
\cite{borland}. When the NFP equation (\ref{pld1}) is coupled to the
Poisson equation (\ref{dyn6}), we obtain the polytropic Smoluchowski
Poisson system describing self-gravitating Langevin particles. When
the NFP equation (\ref{pld1}) is coupled to the field
Eq. (\ref{dyn4}), we obtain a generalized Keller-Segel model of
chemotaxis taking into account anomalous diffusion. These models have
been introduced and studied by Chavanis \& Sire
\cite{lang,csbio}. For the particular index $n_{3}=d/(d-2)$ or $q_{4/3}=\gamma_{4/3}=2(d-1)/d$, the GKS model presents a critical dynamics \cite{csmass}.

\subsection{Generalized Keller-Segel model with logarithmic diffusion: logotropes}
\label{sec_log}

If we take $h(\rho)=1/\rho$ and $g(\rho)=\rho$, we obtain a GKS model
with a logarithmic diffusion
\begin{eqnarray}
\label{log1}
\frac{\partial\rho}{\partial t}=\nabla\cdot \left ( D \nabla\ln\rho-
\chi \rho \nabla c\right ).
\end{eqnarray}
The generalized entropy associated to Eq. (\ref{log1}) is the log-entropy
\begin{eqnarray}
\label{log2}
S=\int \ln\rho \, d{\bf r},
\end{eqnarray}
and the stationary solution is 
\begin{eqnarray}
\label{log3}
\rho=\frac{1}{\alpha-\beta c}. 
\end{eqnarray}
The pressure law is
\begin{eqnarray}
\label{log4}
p(\rho)=T\ln\rho.
\end{eqnarray}
This is similar to a logotropic equation of state \cite{pudritz}. This
is also connected to a polytropic equation of state (or Tsallis
distribution) with $\gamma=q=0$. Indeed, the logotropic model
(\ref{log1}) can be deduced from Eq. (\ref{pld1}) by writing $D\nabla\rho^{q}=Dq\rho^{q-1}\nabla\rho$, taking $q=0$ and
re-defining $Dq\rightarrow D$. When the NFP equation (\ref{log1}) is
coupled to the Poisson equation (\ref{dyn6}), we obtain the logotropic
Smoluchowski-Poisson system. When the NFP equation (\ref{log1}) is coupled to
the field Eq. (\ref{dyn4}), we obtain a generalized Keller-Segel model of
chemotaxis. These models have been introduced and studied by Chavanis
\& Sire \cite{logotrope}.

\subsection{Generalized Keller-Segel models with power law diffusion and power law drift: Tsallis entropy}
\label{sec_pldd}

We introduce here an extension of the GKS  model
(\ref{pld1}). If we take $h(\rho)=q\rho^{q+\mu-1}$ and
$g(\rho)=\rho^{\mu+1}$, we obtain
\begin{eqnarray}
\label{pldd1}
\frac{\partial\rho}{\partial t}=\nabla\cdot \left ( Dq\rho^{q+\mu-1} 
\nabla\rho-
\chi \rho^{\mu+1} \nabla c\right ).
\end{eqnarray}
This corresponds to a power law diffusion
$D(\rho)=\frac{Dq}{q+\mu}\rho^{q+\mu-1}$ and a power law mobility
$\chi(\rho)=\chi \rho^{\mu}$.  The associated 
stochastic process is
\begin{eqnarray}
\label{pldd2}
\frac{d{\bf r}}{dt}=\chi \rho^{\mu}\nabla c+\sqrt{\frac{2Dq}{q+\mu}}\rho^{\frac{q+\mu-1}{2}}{\bf R}(t).
\end{eqnarray}
Since $\rho^{\mu}$ can be put in factor of the diffusion current in
Eq. (\ref{pldd1}), this model has the same equilibrium states
(\ref{pld4}) and the same entropy (\ref{pld3}) as Eq. (\ref{pld1}).

For $\mu=0$, we recover Eq. (\ref{pld1}) with a constant mobility and a power
law diffusion. For $(\mu,q)=(0,0)$, we recover the logotropic
Smoluchowski equation (\ref{log1}) provided that we make the transformation
$Dq\rightarrow D$. For $\mu=1-q$, we have a normal diffusion and a
power law mobility
\begin{eqnarray}
\label{pldd3}
\frac{\partial\rho}{\partial t}=\nabla\cdot \left ( Dq 
\nabla\rho-
\chi \rho^{2-q} \nabla c\right ).
\end{eqnarray}
For $q=2$, we get
\begin{eqnarray}
\label{pldd4}
\frac{\partial\rho}{\partial t}=\nabla\cdot \left (2 D 
\nabla\rho-
\chi  \nabla c\right ),
\end{eqnarray}
which has the same entropy and the same equilibrium states as
Eq. (\ref{pld10}). Finally, for $q=0$ (making the transformation
$qD\rightarrow D$), we obtain
\begin{eqnarray}
\label{pldd6}
\frac{\partial\rho}{\partial t}=\nabla\cdot \left ( D
\nabla\rho-
\chi \rho^{2} \nabla c\right ),
\end{eqnarray}
which has the same entropy and the same equilibrium states as
Eq. (\ref{log1}).  When the NFP equation (\ref{pldd1}) is coupled to
the field equation (\ref{dyn4}), we obtain a generalized Keller-Segel
model of chemotaxis taking into account anomalous diffusion and
anomalous mobility.

\subsection{Generalized Keller-Segel models with a filling factor: Fermi-Dirac entropy}
\label{sec_ff}

If we take $h(\rho)=1$ and $g(\rho)=\rho(1-\rho/\sigma_{0})$, we obtain a GKS model of the form
\begin{eqnarray}
\label{ff1}
\frac{\partial\rho}{\partial t}=\nabla\cdot \left ( D \nabla\rho-
\chi \rho(1-\rho/\sigma_{0}) \nabla c\right ).
\end{eqnarray}
This corresponds to a normal diffusion $D(\rho)=D$ and a mobility
$\chi(\rho)=\chi(1-\rho/\sigma_{0})$ vanishing linearly when the
density reaches the maximum value $\rho_{max}=\sigma_{0}$. The
associated stochastic process is
\begin{eqnarray}
\label{ff2}
\frac{d{\bf r}}{dt}=\chi(1-\rho/\sigma_{0})\nabla c+\sqrt{2D}{\bf R}(t).
\end{eqnarray}
The generalized entropy associated with Eq. (\ref{ff1}) is a
Fermi-Dirac-like entropy in physical space
\begin{eqnarray}
\label{ff3}
S=-\sigma_{0}\int \left\lbrace \frac{\rho}{\sigma_{0}}\ln\frac{\rho}{\sigma_{0}}+\left (1-\frac{\rho}{\sigma_{0}}\right)\ln \left (1-\frac{\rho}{\sigma_{0}}\right)\right\rbrace  d{\bf r},
\end{eqnarray}
and the stationary solution is a Fermi-Dirac-like distribution in
physical space
\begin{eqnarray}
\label{ff4}
\rho=\frac{\sigma_{0}}{1+ e^{-\beta c+\alpha}}.
\end{eqnarray}
From Eq. (\ref{ff4}), we see that, in the stationary state,
$\rho<\sigma_{0}$. This bound is similar to the Pauli exclusion
principle in quantum mechanics. In fact, we can show that $\rho({\bf
r},t)$ remains bounded by $\sigma_{0}$ during the whole evolution.
For $\sigma_{0}\rightarrow +\infty$, we recover the standard KS model
(\ref{sm1}).

An alternative GKS model, with the same entropy and the same equilibrium
states, is obtained by taking $h(\rho)=1/(1-\rho/\sigma_{0})$ and
$g(\rho)=\rho$. This leads to
\begin{eqnarray}
\label{ff5}
\frac{\partial\rho}{\partial t}=\nabla\cdot \left (-D\sigma_{0} \nabla\ln(1-\rho/\sigma_{0})-
\chi \rho \nabla c\right ).
\end{eqnarray}
This corresponds to a nonlinear diffusion with $D(\rho)=-
\sigma_{0}(D/\rho)\ln (1-\rho/\sigma_{0})$ and a constant mobility
$\chi(\rho)=\chi$. Equation (\ref{ff5}) can be put in the form of a
generalized Smoluchowski equation (\ref{gs1})  with a pressure law
\begin{eqnarray}
\label{ff6}
p(\rho)=-T\sigma_{0} \ln(1-\rho/\sigma_{0}).
\end{eqnarray}
For $\rho\ll\sigma_{0}$, we recover the ``isothermal'' equation of
state $p=\rho T$ leading to the standard Keller-Segel model
(\ref{sm1}). However, for higher densities, the equation of state is
modified and the pressure diverges when $\rho\rightarrow
\sigma_{0}$. This prevents the density from exceeding the maximum
value $\sigma_{0}$.

The NFP equation (\ref{ff1}) has been introduced
by Kaniadakis \& Quarati \cite{kq} to describe fermionic systems and by  Robert
\& Sommeria \cite{rs}  in the statistical mechanics of
two-dimensional turbulence (see also \cite{csr}). In the context of
chemotaxis, the model (\ref{ff1}) has been introduced by Hillen \&
Painter \cite{hillen} and, independently, by Chavanis
\cite{gen,degrad}.  It provides a regularization of the standard
Keller-Segel model preventing overcrowding, blow-up and unphysical
singularities. The filling factor $(1-\rho/\sigma_{0})$ takes into
account the fact that the particles cannot interpenetrate because of
their finite size $a$. Therefore, the maximum allowable density is
$\sigma_{0}\sim 1/a^{d}$. It is achieved when all the cells are packed
together. In the model (\ref{ff1}), it is assumed that the mobility vanishes
when the density reaches the close packing value ($\rho\rightarrow
\sigma_{0}$) while the diffusion is not affected. The alternative model (\ref{ff5}) has been introduced by Chavanis \cite{gen,degrad}. 
In that case, the mobility is assumed to be constant and the
regularization preventing overcrowding is taken into account in the
pressure law (\ref{ff6}). We can also multiply the diffusion term and
the mobility term in the NFP equation (\ref{dyn3}) by the {\it same}
positive function $\lambda({\bf r},t)$ in order to obtain a more
general model with the same entropy and the same equilibrium states
in which both diffusion and mobility are affected by overcrowding.

\subsection{Generalized Keller-Segel models incorporating anomalous diffusion and filling factor}
\label{sec_mixed}

The previous models focus individually on two important effects:
anomalous diffusion (see Secs. \ref{sec_pld}-\ref{sec_pldd}) and
exclusion constraints when the density becomes too large (see
Sec. \ref{sec_ff}). Here we introduce a mixed model which combines
these two effects in a single equation. If we take
$h(\rho)=q\rho^{q+\mu-1}$ and
$g(\rho)=\rho^{\mu+1}(1-\rho/\sigma_{0})$, we obtain
\begin{eqnarray}
\label{mixed1}
\frac{\partial\rho}{\partial t}=\nabla\cdot \left ( Dq \rho^{q+\mu-1}\nabla\rho-\chi \rho^{\mu+1}(1-\rho/\sigma_{0}) \nabla c\right ).
\end{eqnarray}
This corresponds to a power law diffusion such that
$D(\rho)=\lbrack{Dq}/({q+\mu})\rbrack \rho^{q+\mu-1}$ and a mobility
$\chi(\rho)=\chi\rho^{\mu}(1-\rho/\sigma_{0})$.  The associated
stochastic process is
\begin{eqnarray}
\label{mixed2}
\frac{d{\bf r}}{dt}=\chi \rho^{\mu}(1-\rho/\sigma_{0})\nabla c+\sqrt{\frac{2Dq}{q+\mu}}\rho^{\frac{q+\mu-1}{2}}{\bf R}(t).
\end{eqnarray}
The generalized entropy corresponding to Eq. (\ref{mixed1}) is obtained by integrating twice the relation 
\begin{eqnarray}
\label{mixed3}
C''(\rho)=\frac{q\rho^{q-2}}{1-\rho/\sigma_{0}}.
\end{eqnarray}
A first integration gives
\begin{eqnarray}
\label{mixed4}
C'(\rho)=q\sigma_{0}^{q-1}\Phi_{q-2}\left(\frac{\rho}{\sigma_{0}}\right ),
\end{eqnarray}
where 
\begin{eqnarray}
\label{mixed5}
\Phi_{m}(t)=\int_{0}^{t}\frac{x^{m}}{1-x}dx.
\end{eqnarray}
Therefore, the generalized entropy can be expressed as
\begin{eqnarray}
\label{mixed6}
C(\rho)=q\sigma_{0}^{q}\int_{0}^{\rho/\sigma_{0}}\Phi_{q-2}(t)dt.
\end{eqnarray}
On the other hand, the equilibrium density is given by
$\rho=\sigma_{0}\Phi^{-1}_{q-2}\lbrack (\beta
c-\alpha)/q\sigma_{0}^{q-1}\rbrack$. Note that these results not
depend on $\mu$ since the term $\rho^{\mu}$ can be put in factor of
the diffusion current in Eq. (\ref{mixed1}).

Let us consider some particular cases. (i) For $q=1$,
Eq. (\ref{mixed1}) has the same entropy and the same equilibrium
states as Eq. (\ref{ff1}). (ii) For $\sigma_{0}\rightarrow +\infty$,
we recover Eq. (\ref{pldd1}). (iii) For $\mu=0$ and $q=2$, we have
\begin{eqnarray}
\label{mixed7}
\frac{\partial\rho}{\partial t}=\nabla\cdot \left ( D \nabla\rho^{2}-
\chi \rho (1-\rho/\sigma_{0}) \nabla c\right ).
\end{eqnarray}
The generalized entropy is 
\begin{eqnarray}
\label{mixed8}
S=-2\sigma_{0}^{2}\int  \left (1-\frac{\rho}{\sigma_{0}}\right)\ln \left (1-\frac{\rho}{\sigma_{0}}\right) d{\bf r},
\end{eqnarray}
and the stationary solution is 
\begin{eqnarray}
\label{mixed9}
\rho=\sigma_{0}\left\lbrack 1-e^{(-\beta c+\alpha)/2\sigma_{0}}\right\rbrack_{+}.
\end{eqnarray}
For $\sigma_{0}\rightarrow +\infty$, we recover Eq. (\ref{pld12}).
Dividing the diffusion and the drift term by $1-\rho/\sigma_0$, we can
also consider the alternative model
\begin{eqnarray}
\label{mixed10}
\frac{\partial\rho}{\partial t}=\nabla\cdot \left (\frac{2\rho D}{1-\rho/\sigma_{0}}\nabla\rho-
\chi \rho  \nabla c\right ),
\end{eqnarray}
which has the same entropy and the same equilibrium states as Eq. (\ref{mixed7}). The pressure law is
\begin{eqnarray}
\label{mixed11}
p(\rho)=-2T\sigma_{0}^{2}\left\lbrack \ln(1-\rho/\sigma_{0})-\rho/\sigma_{0}\right \rbrack.
\end{eqnarray}
(iv) For $(\mu,q)=(0,0)$ and performing the transformation
$qD\rightarrow D$, or directly taking $h(\rho)=1/\rho$ and
$g(\rho)=\rho(1-\rho/\sigma_{0})$, we  obtain
\begin{eqnarray}
\label{mixed12}
\frac{\partial\rho}{\partial t}=\nabla\cdot \left ( D \nabla\ln\rho-
\chi \rho (1-\rho/\sigma_{0}) \nabla c\right ).
\end{eqnarray}
This corresponds to a logarithmic diffusion and a modified mobility
taking into account an exclusion principle through the filling factor. The generalized entropy is obtained from the relation
\begin{eqnarray}
\label{mixed13}
C''(\rho)=\frac{1}{\rho^{2}(1-\rho/\sigma_{0})}, 
\end{eqnarray}
leading to
\begin{eqnarray}
\label{mixed14}
C'(\rho)=-\frac{1}{\sigma_{0}}\left\lbrace \ln\left (\frac{\sigma_{0}}{\rho}-1\right )+\frac{\sigma_{0}}{\rho}\right\rbrace,
\end{eqnarray}
and finally to the explicit expression
\begin{eqnarray}
\label{mixed15}
S=-\int \left (1-\frac{\rho}{\sigma_{0}}\right )\ln \left (\frac{\sigma_{0}}{\rho}-1\right ) d{\bf r}.
\end{eqnarray}
We can consider the alternative model 
\begin{eqnarray}
\label{mixed16}
\frac{\partial\rho}{\partial t}=\nabla\cdot \left \lbrack \frac{D}{\rho(1-\rho/\sigma_{0})}\nabla\rho-
\chi \rho  \nabla c\right \rbrack,
\end{eqnarray}
with the same entropy and the same equilibrium states. The associated pressure law is
\begin{eqnarray}
\label{mixed17}
p(\rho)=-T\ln\left (\frac{\sigma_{0}}{\rho}-1\right ).
\end{eqnarray}

\section{Conclusion}
\label{sec_conc}

In this paper, we have discussed a generalized class of Keller-Segel
models describing the chemotaxis of biological populations. We have
shown their analogy with nonlinear mean field Fokker-Planck equations
and generalized thermodynamics.  We have given explicit examples
corresponding to different entropy functionals. In particular, we have
considered the case where the particles (cells) experience anomalous
diffusion and the case where they experience an exclusion constraint
(volume filling). We have introduced a mixed model taking into account
these two effects in a single equation (\ref{mixed1}). Of course, we
can construct other types of Keller-Segel models which may also be of
interest. The general study of these models, which combine both
long-range interactions and generalized thermodynamics, is
very rich and can lead to a wide diversity of phase transitions and
blow up phenomena. These nonlinear meanfield Fokker-Planck equations
are therefore of considerable theoretical interest \cite{gen}.

\end{document}